%% file: interferenceDecoding-transit-arxiv28Jan2011.tex
\documentclass[letterpaper,onecolumn]{IEEEtranBB}   

\include{defns}

\usepackage{times}
\usepackage[T1]{fontenc}
\usepackage{graphicx}
\graphicspath{{figs/}}

\usepackage[cmex10]{amsmath}
\usepackage{amssymb,textcomp,gensymb} 

\usepackage{float}
\usepackage{accents}

\usepackage[tight,small,bf]{subfigure}   

\usepackage{microtype}

\usepackage{caption}
\DeclareCaptionLabelSeparator{periodquad}{. \quad}
\usepackage{anysize}
	\marginsize{4cm}{4cm}{2cm}{2cm}  

\usepackage[colorlinks=true,linkcolor=black,anchorcolor=black,citecolor=black,filecolor=black,menucolor=black,urlcolor=black]{hyperref}  
 	\hypersetup{
		pdfauthor = {Bernd Bandemer and Abbas El Gamal},
		pdftitle = {Interference Decoding for Deterministic Channels},
		pdfsubject = {Information theory},
		pdfkeywords = {capacity region, deterministic model, interference alignment, interference channel, multiuser information theory, network information theory, simultaneous non-unique decoding},
		pdfcreator = {LaTeX}
	}
\usepackage{cite}  

\renewcommand{\eps}{\varepsilon}

\newcommand{\annleq}[1]{\overset{\text{(#1)}}{\leq}}  
\newcommand{\binSet}{\mathbb F_2}

\newcommand{\dhat}[1]{\ring{#1}}
\newcommand{\Typ}{\mathcal T_\varepsilon^{(n)}}
\newcommand{\Er}{\mathcal E}
\newcommand{\cond}{\,|\,}

\newfloat{tabelle}{tbh}{tabelle}
\floatname{tabelle}{Table}

\newtheorem{thm}{Theorem}

\newtheorem{remark}{Remark}

\newtheorem{lemma}{Lemma}

\title{Interference Decoding \\ for Deterministic Channels}

\author{Bernd Bandemer and Abbas El Gamal \\
Information Systems Laboratory, Stanford University,  \\
350 Serra Mall, Stanford, CA 94305, USA \\
Email: bandemer@stanford.edu, abbas@ee.stanford.edu
	\thanks{\hrule \vspace{2mm} \noindent This work is partially supported by DARPA ITMANET. Bernd Bandemer is supported by an Eric and Illeana Benhamou Stanford Graduate Fellowship.}
}

\begin{document}
\maketitle

\begin{abstract}
An inner bound to the capacity region of a class of deterministic interference channels with three user pairs is presented. The key idea is to simultaneously decode the combined interference signal and the intended message at each receiver. It is shown that this interference-decoding inner bound is tight under certain strong interference conditions. The inner bound is also shown to strictly contain the inner bound obtained by treating interference as noise, which includes interference alignment for deterministic channels. The gain comes from judicious analysis of the number of combined interference sequences in different regimes of input distributions and message rates. Finally, the inner bound is generalized to the case where each channel output is observed through a noisy channel.
\end{abstract}

\section{Introduction}
Interference channels with three or more user pairs exhibit the interesting property that decoding at each receiver is impaired by the \emph{joint} effect of interference from all other senders rather than by each sender's signal separately. Consequently, dealing directly with the effect of the \emph{combined interference signal} is expected to achieve higher rates. 

One such coding scheme is \emph{interference alignment}, e.g.,~\cite{MaddahAli2008,CadambeJafar2008}, in which the code is designed so that the combined interference signal at each receiver is confined (\emph{aligned}) to a subset of the receiver signal space. Depending on the specific channel, this alignment may be achieved via linear subspaces, signal scale levels, time delay slots, or number-theoretic bases of rationally independent real numbers~\cite{EtkinOrdentlich09,Motahari09}. 
In some cases, e.g., the multiple-input multiple-output (MIMO) Gaussian interference channel~\cite{CadambeJafar2008}, the decoder simply treats interference as noise. In general, however, decoding can be thought of as a two-step procedure. In the first step, the received signal is projected onto the desired signal subspace, e.g., by multiplying it by a matrix as for the MIMO case~\cite{MaddahAli2008,CadambeJafar2008} or by separating each received symbol into its constituent lattice points as for the scalar Gaussian case~\cite{Motahari09}. In the second step, interference-unaware decoding is performed on the projection of the received signal. This decoding procedure often leads to an implicit decoding of some function of the undesired messages. Explicit decoding of the combined interference signals was first discussed in~\cite{BreslerParekhTse08} for the many-to-one Gaussian interference channel. The authors argue that with Gaussian codes, decoding the combined interference is tantamount to decoding each interfering sender's codeword. On the other hand, with structured (lattice) codes, the combined interference can be made to appear essentially as a codeword from a single interferer.

In general, for channels with inherent linearity such as Gaussian interference channels, it is natural to consider decoding linear combinations of interfering codewords, instead of individual codewords. This idea is developed in~\cite{NazerGastpar09} for Gaussian relay networks, leading to a {\em compute--forward} relaying scheme. 

\begin{figure}[t!]   
	\centering
	\includegraphics{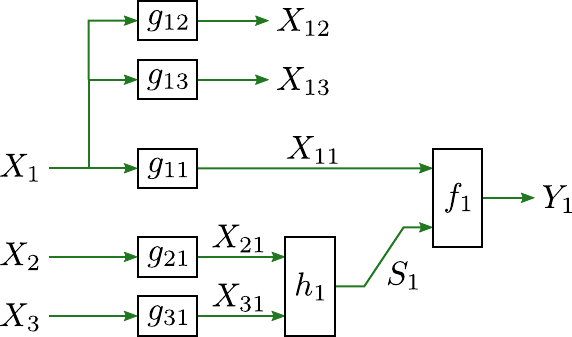}
	\caption{Block diagram of the 3-DIC for the first receiver.}
	\label{fig:gendet3ic}
\end{figure}

In this paper, we investigate interference decoding for the three receiver deterministic interference channel (3-DIC) depicted in Figure~\ref{fig:gendet3ic}. The channel consists of three sender-receiver alphabet pairs $(\Xc_k,\Yc_k)$, loss functions $g_{lk}$, interference combining functions $h_k$, and receiver functions $f_k$ for $k,l\in \{1,2,3\}$. The outputs of the channel are
\begin{align}
Y_{k} &= f_k( X_{kk}, S_{k} ), \quad \text{where} \label{eq:3dicOutput} \\
X_{lk} &= g_{lk}(X_{l}), \notag \\
S_{1} &= h_1(X_{21},X_{31}), \notag \\
S_{2} &= h_2(X_{12},X_{32}), \notag  \\
S_{3} &= h_3(X_{13},X_{23}). \notag 
\end{align}
We assume that $h_k$ and $f_k$ are one-to-one when either one of their arguments is fixed. For example, for $Y_1 = f_1(X_{11}, S_1)$, this assumption is equivalent to $H(X_{11})=H(Y_1 \cond S_1)$ and $H(S_1)=H(Y_1 \cond X_{11})$ for every probability mass function (pmf) $p(x_{11}, s_1)$. 
Except for requiring the one-to-one property to hold for both arguments, this channel model is an extension of the Costa--El Gamal two-user-pair model~\cite{ElGamalCosta82}. Note that this model is more general than the class of deterministic interference channels studied in~\cite{GouJafar2009}.

Each sender $k \in \{1,2,3\}$ wishes to convey an independent message $M_k$ at data rate $R_k$ to its corresponding receiver. We define a $(2^{nR_1}, 2^{nR_2}, 2^{nR_3},n)$ code, probability of error, and achievability of a given rate triple $(R_1,R_2,R_3)$ in the standard way (see~\cite{ElGamalKimLecture}). 

We focus on this class of deterministic channels for several reasons. First, the capacity region for the two-user-pair version of this class~\cite{ElGamalCosta82} is known and is achieved by the Han--Kobayashi scheme~\cite{HanKobayashi81}. This gives some hope that an appropriate extension of Han--Kobayashi where the combined interference is decoded partially or fully may be optimal for more than two user pairs. Second, this class includes the finite field deterministic model in~\cite{Avestimehr07}, which approximates Gaussian interference channels in the high SNR regime~\cite{BreslerTse08}. For more than two user pairs, capacity results for the finite field deterministic model are known only in some special cases~\cite{JafarVishwanath08, Bandemer09}, where interference is treated as noise. An interesting question is whether more sophisticated coding schemes can achieve higher rates for this class of channels. Finally, the combined interference signal in our channel takes values from a finite set, and therefore a certain type of alignment can be observed without resorting to complicated structured codes~\cite{NazerGastpar2008}. 

The main result of the paper is an inner bound on the capacity region of the 3-DIC, which is achieved via \emph{interference decoding}. We assume point-to-point codes without rate splitting or superposition coding since such codes are widely deployed and it is interesting to investigate the benefit of using a more sophisticated receiver instead of treating interference as noise. Specifically, each receiver simultaneously decodes the intended message and the combined interference without penalizing incorrect decoding of the latter.  Of course, one does not expect this scheme to be optimal in general, since even for the two-user-pair case, superposition coding is required for optimality. Note that for our class of deterministic channels, algebraic structures such as linear subspaces or lattices do not exist in general. Hence, our decoder does not use the two-step procedure as in the work on Gaussian channels and their corresponding high SNR deterministic models.

The key observation is that depending on the input pmfs and the message rates, the number of possible combined interference sequences can be equal to the number of interfering message pairs, the number of typical combined interference sequences, or some combination of the two.  In our scheme, each sender does not need to know the other senders' codebooks. However, we use simultaneous decoding, which requires that the receivers know all codebooks. As in the recent characterization of the Han--Kobayashi region~\cite{Chong08}, we do not require the interference decoding to be correct with arbitrarily small probability of error.

In the following section, we summarize and discuss the main results in this paper. The proofs of these results are given in Sections~\ref{sec:proof_ID},~\ref{sec:proof_ID_is_capacity} and~\ref{sec:proof_inclusion}, with some details deferred to the Appendix. In Section~\ref{sec:finalRemarks}, we give final remarks on the optimality of interference decoding. Throughout the rest of the paper, notation and basic definitions follow~\cite{ElGamalKimLecture}.

\section{Summary of main results}  \label{sec:mainresults}
The main results in this paper are as follows.

\subsection*{Interference-decoding inner bound} 
Fix the random tuple $(Q,X_1,X_2,X_3)\sim p(q)p(x_1|q)$ $p(x_2|q)p(x_3|q)$, where $Q$ is a time-sharing random variable from alphabet $\Qc$. Define the region $\Rr_1(Q,X_1,X_2,X_3)$ to consist of the rate triples $(R_1,R_2,R_3)$ such that
	\begin{align}
			R_1 &< H(X_{11} \cond Q), \label{eq:IDcond1}  \\
			R_1 + \min\{ R_2, H(X_{21}\cond Q) \} &< H(Y_1\cond  X_{31},Q), \label{eq:IDcond2} \\
			R_1 + \min\{ R_3, H(X_{31}\cond Q) \} &< H(Y_1\cond  X_{21},Q), \label{eq:IDcond3} \\
			R_1 + \min\{ R_2+R_3, \ \, \notag \\
				R_2+H(X_{31}\cond Q), \ \, \notag  \\
				H(X_{21}\cond Q)+R_3, \ \, \notag \\
				H(S_1\cond Q)
				\} &<  H(Y_1\cond Q). \label{eq:IDcond4}  
	\end{align}
Similarly define the regions $\Rr_2(Q,X_1,X_2,X_3)$ and $\Rr_3(Q,X_1,X_2,X_3)$ by making the subscript replacements $1\mapsto 2 \mapsto 3 \mapsto 1$  and $1\mapsto 3 \mapsto 2 \mapsto 1$ in $\Rr_1(Q,X_1,X_2,X_3)$, respectively. 
\begin{thm}[Interference-decoding inner bound] \label{thm:ID} 
The region
\[
\Rr_{\rm ID}=\bigcup_{(Q,X_1,X_2,X_3)} \bigcap_{k=1}^3 \Rr_k(Q,X_1,X_2,X_3),
\] 
where $(Q,X_1,X_2,X_3)\sim p(q)p(x_1|q)p(x_2|q)p(x_3|q)$ and $|\Qc| \le 13$ is an inner bound to the capacity region of the 3-DIC. 
\end{thm}
The proof for this theorem is given in Section~\ref{sec:proof_ID}.

Region $\Rr_k(Q,X_1,X_2,X_3)$ ensures decodability at receiver $k$. The $\min$ terms on the left hand side of the inequalities arise from counting the number of possible interfering sequences at various links of the channel. For example, consider the $\min\{ R_2, H(X_{21}\cond Q) \}$ term in~\eqref{eq:IDcond2}. If $R_2$ is small, the number of distinct sequences that can occur at $X_{21}$ is equal to the number of possible messages from sender 2. As $R_2$ increases beyond $H(X_{21}\cond Q)$, the the number of possible sequences at $X_{21}$ ``saturates'' to the number of typical sequences, which is roughly $2^{nH(X_{21}\cond Q)}$. In this case, we can increase the rate of the second sender further without negatively  impacting the first receiver. The $\min$ expressions in~\eqref{eq:IDcond3} and~\eqref{eq:IDcond4} likewise capture the saturation effects at $X_{31}$ and $S_1$, respectively. 

\begin{figure}[t!]
\centering
\includegraphics{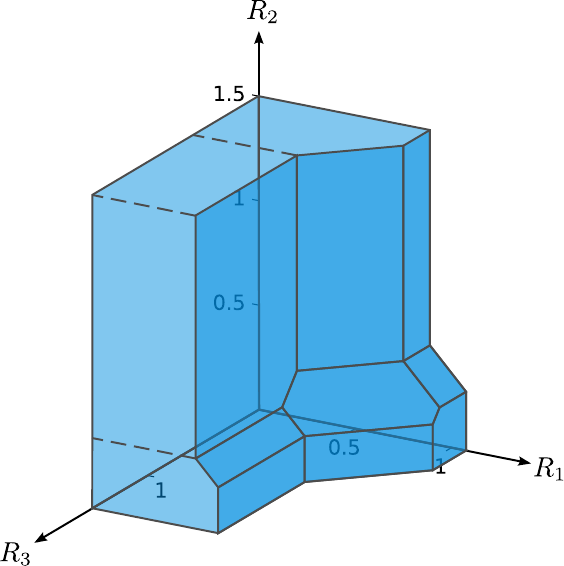}
\caption{Region $\Rr_1$ showing allowable rate triples for decodability at the first receiver.}
\label{fig:rot2}
\end{figure}

An example of region $\Rr_1(Q,X_1,X_2,X_3)$ is plotted in Figure~\ref{fig:rot2}. The region is unbounded in the $R_2$ and $R_3$ directions, due to saturation. This is expected, since regardless of the values of $R_2$ and $R_3$, $S_1$ can always be treated as noise to achieve a non-zero rate. However, as $R_2$ and $R_3$ become smaller, the proposed scheme takes advantage of the structure in $S_1$ and can thereby increase $R_1$. 

\subsection*{Capacity region under strong interference} 
Consider the subclass of 3-DIC \emph{with strong interference and invertible $h_k$} in which the following two conditions hold. 

First, the loss functions $g_{lk}$ are such that
\begin{align*}
	\min\{ H(X_{12}), H(X_{13}) \} &\geq H(X_{11}), \\
	\min\{ H(X_{21}), H(X_{23}) \} &\geq H(X_{22}), \\
	\min\{ H(X_{31}), H(X_{32}) \} &\geq H(X_{33}), 
\end{align*}
for all product input pmfs $p(x_1)p(x_2)p(x_3)$. This condition implies that interference is strong. 

Second, the functions $h_k$ are invertible, i.e.,
\begin{align*}
	H(S_1) &= H(X_{21}) + H(X_{31}), \\
	H(S_2) &= H(X_{12}) + H(X_{32}), \\
	H(S_3) &= H(X_{13}) + H(X_{23}),
\end{align*}
for all product input pmfs $p(x_1)p(x_2)p(x_3)$. With the conditional invertibility property of $f_k$, the channel becomes a non-symmetric version of the deterministic model for the SIMO interference channel described in~\cite{GouJafar2009}. In both cases, a receiver can uniquely recover \emph{both} interfering signals given the received sequence and the desired transmitted sequence. The capacity region under these conditions is achieved by interference decoding. 

\begin{thm} \label{thm:ID_is_capacity} 
 The capacity region of the 3-DIC under strong interference and invertible $h_k$ functions is the set of rate triples $(R_1,R_2,R_3)$ such that
	\begin{align*}
				R_k &< H(X_{kk} \cond Q), \quad k \in \{1,2,3\},      \\
				R_1+R_2 &< \min \{ H(Y_1\cond X_{31},Q), H(Y_2 \cond X_{32}, Q) \}, \\
				R_1+R_3 &< \min \{ H(Y_1\cond X_{21},Q), H(Y_3 \cond X_{23}, Q) \}, \\
				R_2+R_3 &< \min \{ H(Y_2\cond X_{12},Q), H(Y_3 \cond X_{13}, Q) \}, \\
				R_1+R_2+R_3 &< \min \{ H(Y_1\cond Q), H(Y_2\cond Q), H(Y_3\cond Q)  \} ,
	\end{align*}
for some $(Q,X_1,X_2,X_3)\sim p(q)p(x_1|q)p(x_2|q)p(x_3|q)$. 
\end{thm}
The proof of this theorem is given in Section~\ref{sec:proof_ID_is_capacity}.

\subsection*{Treating interference as noise}
In the two-user-pair interference channel, decoding both messages at each receiver and treating interference as noise are considered as two extreme schemes. The extremes are bridged by the Han--Kobayashi scheme in which part of the interference is decoded and the rest is treated as noise~\cite{ElGamalKimLecture}. While treating interference as noise is better for channels with weak interference, decoding both messages is optimal under strong interference. We show surprisingly that for the 3-DIC under consideration, treating interference as noise is a \emph{special case} of interference decoding! 

By using randomly and independently generated codebooks as for the interference-decoding inner bound, but having each receiver decode only its message, we obtain the following inner bound. 
\begin{lemma}[Treating interference as noise] \label{lemma:tian}
	The set $\Rr_{\rm TIN}$ of rate triples $(R_1,R_2,R_3)$ such that
	\begin{align} \label{eq:tian}
		R_k &<  I(X_k; Y_k\cond Q), \quad k \in \{1,2,3\},
	\end{align}
	for some pmf $p(q)p(x_1|q)p(x_2|q)p(x_3|q)$ constitutes an inner bound to the capacity region of the general three-user-pair memoryless interference channel. 
\end{lemma}
Note that in contrast to interference decoding, a user pair does not need to know the codebooks of other user pairs. Also note that this inner bound, with appropriate selections of the input pmfs, includes the interference alignment inner bounds in~\cite{CadambeJafar2008,JafarVishwanath08,Bandemer09}.   Maximum alignment is achieved when the number of combined interference sequences, e.g., $S_1^n$, is much smaller than the number of individual interference sequence pairs, e.g., $(X_{21}^n, X_{31}^n)$. Since $I(X_k; Y_k\cond Q) = H(Y_k\cond Q) - H(S_k\cond Q)$, this occurs when $H(S_k\cond Q)$ is small, causing the number of $S_k^n$ sequences to saturate. 

In Section~\ref{sec:proof_inclusion}, we establish the following result.
\begin{thm} \label{thm:inclusion}
	The rate region achievable by treating interference as noise is included in the interference-decoding rate region, i.e., $\Rr_{\rm TIN} \subseteq \Rr_{\rm ID}$.
\end{thm}

The difference between treating interference as noise and interference decoding is essentially that the former assumes that the combined interference signal $S_k$ is always saturated, while the latter distinguishes between saturated and non-saturated cases. Later in this section, we argue that the above  inclusion result is tightly coupled to the definition of the 3-DIC.

The following example shows that the inclusion of Theorem~\ref{thm:inclusion} can be strict, i.e., the treating interference as noise region is strictly contained in the interference-decoding region.  

\subsubsection*{Additive 3-DIC example} 
\begin{figure}[t!]
	\centering
	\includegraphics{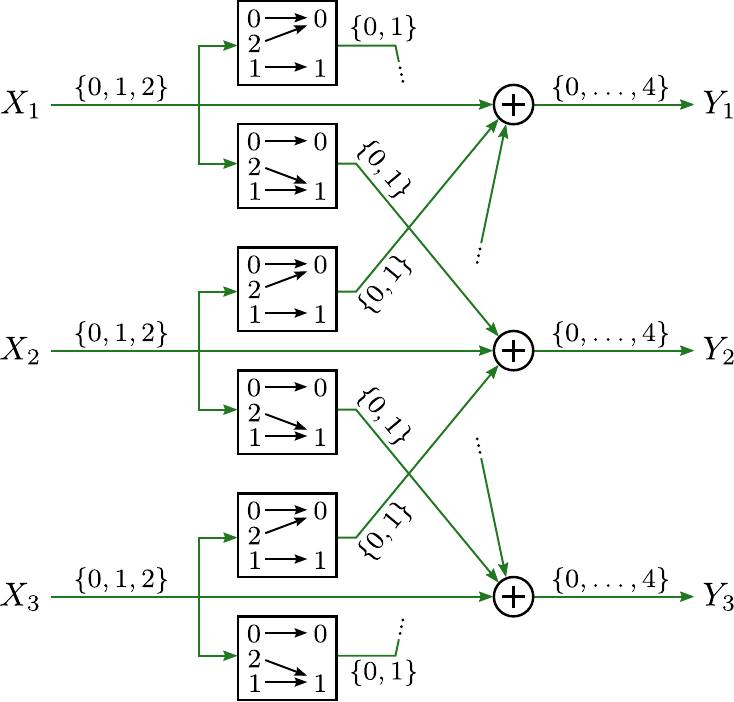}
	\caption{Additive 3-DIC example.}
	\label{fig:example3dic}
\end{figure}

Consider a cyclically symmetric 3-DIC with $\mathcal X_1=\mathcal X_2=\mathcal X_3=\{0,1,2\}$, and $\mathcal Y_1=\mathcal Y_2=\mathcal Y_3=\{0,1,2,3,4\}$, where $g_{11}=g_{22}=g_{33}=g$, $g_{12}=g_{23}=g_{31}=g_{+}$ and $g_{21}=g_{32}=g_{13}=g_-$ as well as $h_1=h_2=h_3=h$ and $f_1=f_2=f_3=f$. The direct path loss functions are the identity mapping, $g=\mathrm{Id}$, while the cross path loss functions are given by $g_- = \{0 \mapsto 0, 1 \mapsto 1, 2 \mapsto 0\}$ and $g_+ = \{0 \mapsto 0, 1 \mapsto 1, 2 \mapsto 1\}$ (similar to the Blackwell broadcast channel~\cite{vanderMeulen77}). Finally, the combining functions $h$ and receiver functions $f$ are taken to be addition. The resulting input-to-output mapping is shown in Figure~\ref{fig:example3dic}.

\begin{figure}[t!]
	\centering
	\subfigure[Treating interference as noise]{
		\includegraphics{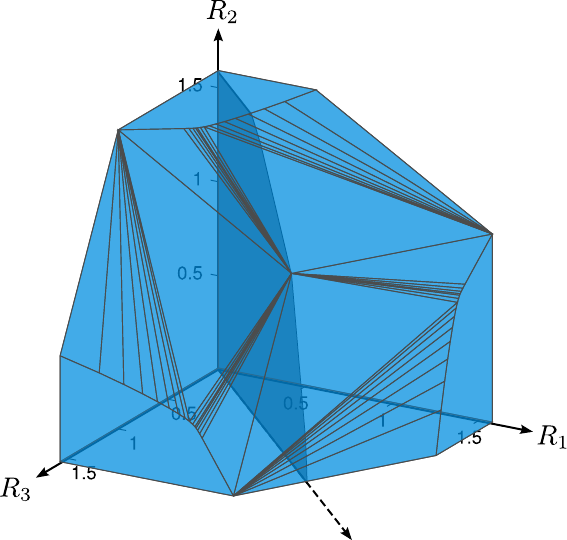}
		\label{fig:IntfAsNoise}
	}
	\hfill
	\subfigure[Interference decoding]{
		\includegraphics{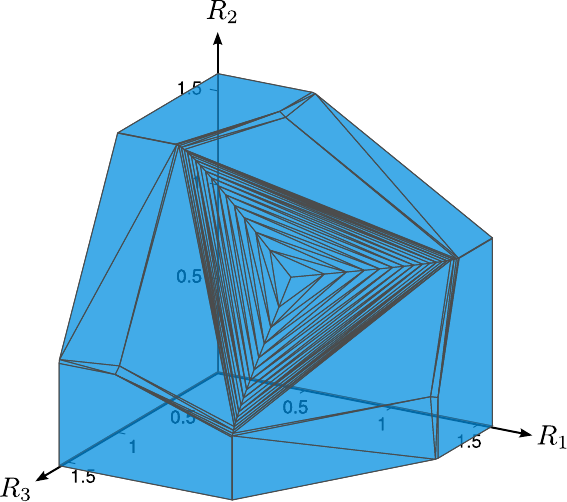}
		\label{fig:IntfDecode}
	}
	\hfill
	\subfigure[Intersection with 45\degree-plane]{
		\includegraphics{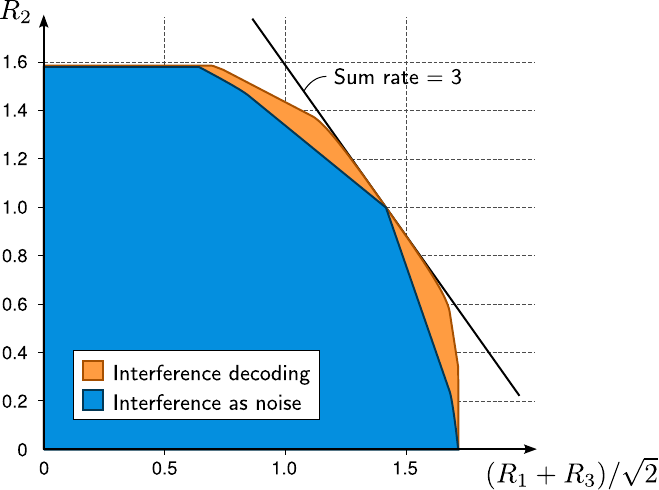}
		\label{fig:2dCut}
	}
	\caption{Inner bounds for the additive 3-DIC example. }
	\label{fig:example2plots}
\end{figure}

For this channel, the interference-decoding rate region strictly contains the region achievable by treating interference as noise. To demonstrate this, we computed the approximations of the inner bounds shown in Figures~\ref{fig:IntfAsNoise} and~\ref{fig:IntfDecode}. Since it is computationally infeasible to enumerate the $13$ different conditional distributions of inputs given $Q$ as required by Theorem~\ref{thm:ID}, we used the following procedure. We first assume $Q=\emptyset$ and consider a grid over all input distributions $p(x_1)p(x_2)p(x_3)$. For each grid point, we compute the achievable rate regions as given by Theorem~\ref{thm:ID} and Lemma~\ref{lemma:tian}, respectively. We represent the regions as the convex hull of its corner points. The final approximation is obtained by taking the union of all such corner points over the grid. Due to the simple structure of $\Rr_{\rm TIN}$ in Lemma~\ref{lemma:tian}, which consists of a union of rectangular boxes, this method can compute $\Rr_{\rm TIN}$ to arbitrary precision provided the grid is sufficiently fine. On the other hand, our approximation method yields a possibly strictly smaller inner bound than $\Rr_{\rm ID}$. 

Figure~\ref{fig:2dCut} depicts the intersection of the three-dimensional regions with the plane defined by the $R_2$ axis and the 45$\degree$-line between the $R_1$ and $R_3$ axes. This plane is also shown in Figure~\ref{fig:IntfAsNoise}.
Note that the same maximum sum rate $R_\text{sum}=3$ is achieved by both schemes. However, while treating interference as noise does so at exactly one rate triple ($R_1=R_2=R_3=1$), interference decoding achieves the maximal sum rate at many different asymmetric rate triples.

\begin{remark}
Treating interference as noise can be optimal in some cases. 
Consider the three-user-pair cyclically symmetric finite field deterministic model investigated in~\cite{Bandemer09}, which is a special case of the channel considered in this paper. The input and output alphabets for this channel are $\binSet^N$ and $\binSet^{2N}$, respectively, the loss functions $g_{lk}$ are vector shifting operations, where the amount of shift is parameterized by $(\alpha,\beta) \in [1,2]\times[0,1]$, and the interference combining functions $h_k$ and the receiver functions $f_k$ are componentwise additions over $\binSet$. 

The sum capacity of this channel is computed in~\cite{Bandemer09} for a large range of $(\alpha,\beta)$, 
and achievability is established by constructing linear encoding and decoding matrices for every $(\alpha,\beta)$. This scheme can be interpreted as treating interference as noise, and thus Lemma~\ref{lemma:tian} subsumes the achievability results in~\cite{Bandemer09}. In fact, the necessary input distributions are the ones implicitly stated there. 

It would be interesting to investigate whether interference decoding can achieve higher sum rates than treating interference as noise in the $(\alpha,\beta)$ range where the sum capacity is not known. Moreover, even in the range where we know the sum capacity, interference decoding may achieve higher asymmetric rates than treating interference as noise, as in the additive 3-DIC example. The main challenge in settling these questions is the prohibitively large space of possible input distributions in Theorem~\ref{thm:ID}. 
\end{remark}

\subsection*{Extension to 3-DIC with noisy observations}
Finally, we consider the 3-DIC \emph{with noisy observations}. In this generalization of 3-DIC, the channel outputs in~\eqref{eq:3dicOutput} are observed through memoryless channels $Y_k \to Z_k$ for $k \in \{1,2,3\}$. Thus receiver $k$ now observes a noisy version $Z_k$ of $Y_k$, which may be from a discrete or a continuous alphabet. 

The interference-decoding inner bound generalizes to the 3-DIC with noisy observations as follows. Let $(Q,X_1,X_2,X_3)\sim p(q)p(x_1|q)p(x_2|q)p(x_3|q)$. Define the region $\Rr'_1(Q,X_1,X_2,X_3)$ as the set of rate triples $(R_1,R_2,R_3)$ such that
	\begin{align*}
			R_1 &< I(X_1; Z_1 \cond S_1,Q), \\
			R_1 + \min\{ R_2, H(X_{21}\cond Q) \} &< I(X_1, X_{21}; Z_1 \cond X_{31},Q),  \\
			R_1 + \min\{ R_3, H(X_{31}\cond Q) \} &< I(X_1, X_{31}; Z_1 \cond X_{21},Q),  \\
			R_1 + \min\{ R_2+R_3, \ \,   \\
				R_2+H(X_{31}\cond Q), \ \,   \\
				H(X_{21}\cond Q)+R_3, \ \,  \\
				H(S_1\cond Q)
				\} &<  I(X_1,S_1; Z_1 \cond Q).  
	\end{align*}
Similarly, define the regions $\Rr'_2(Q,X_1,X_2,X_3)$ and $\Rr'_3(Q,X_1,X_2,X_3)$ by making the subscript replacements $1\mapsto 2 \mapsto 3 \mapsto 1$  and $1\mapsto 3 \mapsto 2 \mapsto 1$ in $\Rr'_1(Q,X_1,X_2,X_3)$, respectively. 

\begin{thm} \label{thm:ID_noisy} 
The region
\[
\Rr'_{\rm ID}=\bigcup_{(Q,X_1,X_2,X_3)} \bigcap_{k=1}^3 \Rr'_k(Q,X_1,X_2,X_3),
\] 
where $(Q,X_1,X_2,X_3)\sim p(q)p(x_1|q)p(x_2|q)p(x_3|q)$ is an inner bound to the capacity region of the 3-DIC with noisy observations. 
\end{thm}
The proof of this theorem proceeds completely analogously to the proof of Theorem~\ref{thm:ID} as presented in Section~\ref{sec:proof_ID}, and thus its details are omitted.


The following example demonstrates the inner bound for the 3-DIC with noisy observations. It also shows that the inclusion of Theorem~\ref{thm:inclusion} does not hold in general for this channel model.
\subsubsection*{Gaussian interference channel example} 
Consider the Gaussian interference channel with finite input alphabets. The channel output at receiver $k$ is 
\begin{align}
	Y_{k} &= \sum_{l=1}^3 g_{lk} X_{l},  \notag \\
	Z_{k} &= Y_{k} + N_{k},   \label{eq:gaussian_example}
\end{align}
where $g_{lk} \in \Real$ is the path gain from transmitter $l$ to receiver $k$, and $N_{k}$ is additive white Gaussian noise of average power $\sigma^2$. This is a realistic model for a wireless interference channel where the transmitter hardware is based on digital signal processing (DSP) and digital-to-analog conversion (DAC). For example, $\Xc_l=\{+1,-1\}$ represents a system with a binary constellation, e.g., binary phase-shift keying (BPSK). 
Equation~\eqref{eq:gaussian_example} represents continuous-valued outputs (soft outputs), but our model would also apply if a quantizer is added (hard outputs), for example due to analog--digital conversion (ADC) at the receivers. 

\begin{figure}[t!]
	\centering\label{subsec:tian}
	\includegraphics[width=5.5cm]{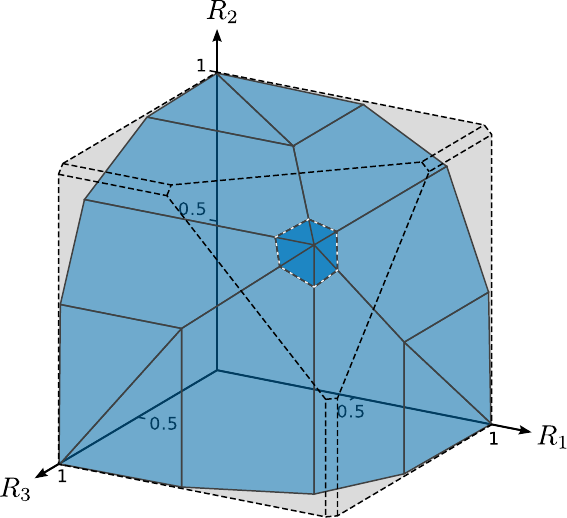}
	\caption{Rate regions achieved by interference decoding (dashed outline) and treating interference as noise (shaded) for a cyclically symmetric Gaussian interference channel with $X_k \in \{+1,-1\}$, path gains $g_{11}=1.8$, $g_{21}=1.0$, $g_{31}=1.1$, and noise power $\sigma^2=0.1$.}
	\label{fig:exampleGaussianBPSK_modified}
\end{figure}

Figure~\ref{fig:exampleGaussianBPSK_modified} shows approximations of the inner bounds for a cyclically symmetric Gaussian interference channel with BPSK inputs and continuous outputs. In contrast to the noiseless case, neither the interference-decoding region nor the region achieved by treating interference as noise contains the other, i.e., Theorem~\ref{thm:inclusion} does not hold for 3-DIC with noisy observations. In particular, the sum rates achieved by treating interference as noise and interference decoding are $2.51$ and $2.37$, respectively. Intuitively, interference decoding attempts to separate the combined interference from the additive noise. As such, it may achieve lower rates than simply treating interference as noise for which this separation is not enforced. This discrepancy is more pronounced for low values of SNR, and it vanishes asymptotically as SNR grows.

\section{Proof of interference-decoding inner bound} \label{sec:proof_ID}
We first present two key lemmas which formalize the notion of link saturation as discussed after Theorem~\ref{thm:ID}. The proofs are deferred to the Appendix. The first lemma generalizes the packing lemma stated in~\cite{ElGamalKimLecture}. 
\begin{lemma}[Packing lemma for pairs] \label{lemma:mpl} Let $(U,A,B,C) \sim p(u) p(a|u) p(b|u)p(c|a,b,u)$. Let $U^n \sim \prod_{i=1}^n p_U(u_i)$. For each $m \in [1:2^{nR_A}]$, let $A^n(m)\sim \prod_{i=1}^n p_{A|U}(a_i\cond u_i)$. For each $l \in [1:2^{nR_B}]$, let $B^n(l)\sim \prod_{i=1}^n p_{B|U}(b_i\cond u_i)$, conditionally independent of each $A^n(m)$ given $U^n$. Let $C^n\sim \prod_{i=1}^n p_{C|U}(c_i\cond u_i)$, conditionally independent of each $A^n(m)$ and $B^n(l)$ given $U^n$. There exists a $\delta(\eps)$ with $\lim_{\eps\to 0} \delta(\eps) = 0$ such that if 
\begin{align*}
	\min\{R_A, H(A\cond U)\} \qquad \\
	+ \min\{R_B, H(B\cond U)\} 
	& < I(A,B;C \cond  U) - \delta(\eps),
\end{align*}
then $\P\{ (U^n, A^n(m), B^n(l), C^n) \in \Typ \text{ for some $m,l$} \} \to 0$ as $n\to \infty$, where 
typicality, entropies and mutual information
are with respect to $p(u,a,b,c)$.
\end{lemma}

The following lemma is a refined version of Lemma~\ref{lemma:mpl}, where the sequences $B^n(l)$ are generated from two conditionally independent components $B_1^n(l_1)$ and $B_2^n(l_2)$. 
\begin{lemma}
\label{lemma:spl} 

Let $(U,A,B_1,B_2,B,C) \sim p(u)p(a|u)p(b_1|u)$ $p(b_2|u)p(b|b_1,b_2)p(c|a,b,u)$, where $p(b | b_1,b_2)$ corresponds to a deterministic mapping $h: (b_1,b_2)\mapsto b$. Let $U^n \sim \prod_{i=1}^n p_U(u_i)$. 
For each $m \in [1:2^{nR_A}]$, let $A^n(m)\sim \prod_{i=1}^n p_{A|U}(a_i\cond u_i)$.
For each $l_1 \in [1:2^{nR_{B_1}}]$, let $B_1^n(l_1)\sim \prod_{i=1}^n p_{B_1|U}(b_{1i}\cond u_i)$, conditionally independent of each $A^n(m)$ given $U^n$. 
Likewise, for each $l_2 \in [1:2^{nR_{B_2}}]$, let $B_2^n(l_2)\sim \prod_{i=1}^n p_{B_2|U}(b_{2i}\cond u_i)$, conditionally independent of each $A^n(m)$ and $B_1^n(l_1)$ given $U^n$. 
For each $(l_1,l_2)$, let $B_i(l_1,l_2) = h( B_{1i}(l_1), B_{2i}(l_2) )$ for $i \in [1:n]$. Finally, let $C^n\sim \prod_{i=1}^n p_{C|U}(c_i\cond u_i)$, conditionally independent of each $A^n(m)$, $B_1^n(l_1)$, and $B_2^n(l_2)$ given $U^n$. 

There exists a function $\delta(\eps)$ with $\lim_{\eps\to 0} \delta(\eps) = 0$ such that if
\begin{align*}
	R_A + \min\{R_{B_1}+R_{B_2}, \ \, \\
	R_{B_1}+H(B_2\cond U), \ \,  \\
	H(B_1\cond U)+R_{B_2}, \ \, \\
	H(B\cond U)\} & < I(A,B;C \cond  U) - \delta(\varepsilon),
\end{align*}
then $\P\{ (U^n, A^n(m), B^n(l_1,l_2), C^n) \in \Typ $ for some $m$, $l_1$, $l_2 \} \to 0$ as $n\to \infty$, where typicality, entropies and mutual information are with respect to $p(u,a,b_1,b_2,b,c)$.
\end{lemma}

\begin{remark} 
	The intuition is that $B$ can be interpreted as the output of a deterministic multiple access channel with inputs $B_1$ and $B_2$ and input to output mapping $h$. Figure~\ref{fig:2mac_nsequences_tall} shows the number of output sequences for different ranges of $R_{B_1}$ and $R_{B_2}$ when $h$ is one-to-one in each argument. Note that when $(R_{B_1},R_{B_2})$ is in the deterministic MAC capacity region, the number of output sequences is simply $2^{n(R_{B_1}+R_{B_2})}$. For $(R_{B_1},R_{B_2})$ outside the capacity region, the number of output sequences saturates in one or both dimensions.   
	The logarithm of the number of output sequences divided by $n$ appears in the $\min$ expression of the lemma.
	\begin{figure}[t!]
		\centering
		\includegraphics{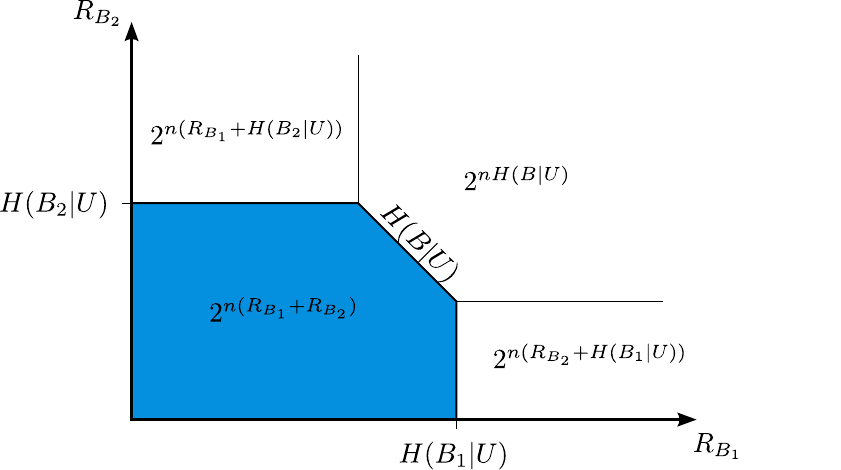}
		\caption{Capacity region for a deterministic MAC. The number of output sequences as a function of the number of input sequences is stated in each region.}
		\label{fig:2mac_nsequences_tall}
	\end{figure}
\end{remark}

We are now ready to prove Theorem~\ref{thm:ID}.  
We begin by fixing a pmf $p(q)p(x_1|q)p(x_2|q)p(x_3|q)$. 

\vspace{2mm}
\noindent \emph{Codebook generation.} \quad Randomly generate a sequence $q^n$ according to $\prod_{i=1}^np_Q(q_i)$. For each $k \in \{1,2,3\}$, randomly and conditionally independently generate sequences $x_k^n(m_k)$, $m_k \in [1:2^{nR_k}]$, each according to $\prod_{i=1}^np_{X_k|Q}(x_{ki}|q_i)$. From the channel definition, this procedure induces intermediate sequences $x_{kl}^n(m_k)$ for $l \in \{1,2,3\}$, combined interference sequences $s^n_1(m_2,m_3)$, $s^n_2(m_1,m_3)$, $s^n_3(m_1,m_2)$, and output sequences $y^n_k(m_1,m_2,m_3)$.

\vspace{2mm}
\noindent \emph{Encoding.} \quad To send the message $m_k \in [1:2^{nR_k}]$, $k \in \{1,2,3\}$, encoder $k$ transmits $x_k^n(m_k)$. 

\vspace{2mm}
\noindent \emph{Decoding.} \quad Upon observing $y_1^n$, decoder 1 declares that $\hat m_1$ is sent if it is the unique message such that $(q^n, x_1^n(\hat m_1),  s^n_1(\dhat m_2, \dhat m_3), x_{21}^n(\dhat m_2), x_{31}^n(\dhat m_3), y_1^n) \in \Typ$ for some $ \dhat m_2, \dhat m_3$, where $\Typ$ is defined as in~\cite{ElGamalKimLecture}.  
Decoding at the other receivers is performed similarly. 

\vspace{2mm}
\noindent \emph{Analysis of the probability of error.}\quad
Without loss of generality, assume that $m_k=1$ for $k \in \{1,2,3\}$. Define $\Er_{mlk} = ( Q^n, X_1^n(m),  S^n_1(l,k), X_{21}^n(l), X_{31}^n(k), Y_1^n(1,1,1)) \in \Typ$, and the events
\begin{align*}
	\Er_0 &= \Er^c_{111}, \\
	\Er_1 &= \left \{ \Er_{m11} \text{ for some } m\neq 1 \right \},\\
	\Er_2 &= \left \{ \Er_{ml1}  \text{ for some } m,l\neq 1\right \},  \\
	\Er_3 &= \left \{ \Er_{m1k}  \text{ for some } m,k\neq 1 \right \},\\
	\Er_4 &= \left \{ \Er_{mlk}  \text{ for some } m,l,k\neq 1 \right \}.
\end{align*}
Then the probability of decoding error at the first receiver averaged over codebooks is upper bounded as $\P(\Er) = \P(\Er_0 \cup \Er_1 \cup \Er_2 \cup \Er_3 \cup \Er_4)\le \sum_{j=0}^4\P(\Er_j)$.
We bound each term. First, by the law of large numbers, $\P(\Er_0) \to 0 $ as $n\to \infty$.  

Next consider
\begin{align*}
	\Er_1 & \subseteq \Bigl \{ \left(Q^n, X_1^n(m), S_1^n(1,1), Y_1^n(1,1,1) \right) 
	 \in \Typ  \\
	 & \qquad \text{ for some } m \neq 1 \Bigr \}.
\end{align*}
By Lemma~\ref{lemma:mpl} with $U^n=(Q^n,S^n_1(1,1))$, $A^n=X_1^n$, $B^n=\emptyset$, and $ C^n=Y^n_1(1,1,1)$, the probability of this event tends to zero as $n \to \infty$ if
\begin{align*}
	R_1 & < I(X_1; Y_1 \cond  S_1,Q),
\end{align*}
which simplifies to
\begin{align}
	R_1 & < H(X_{11} \cond Q). \label{eq:firstcond2}
\end{align}
The event $\Er_2$ can be treated similarly. Consider
\begin{align*}
	\Er_2 & \subseteq \Bigl \{  \left( Q^n, X_1^n(m),  X_{21}^n(l), X_{31}^n(1), Y_1^n(1,1,1) \right) \in \Typ \\
	& \qquad \text{ for some } m,l\neq 1 \Bigr \}.
\end{align*}
Using Lemma~\ref{lemma:mpl} with $U^n=(Q^n,X^n_{31}(1))$, $A^n=X^n_1$, $B^n=X^n_{21}$, and $C^n=Y^n_1(1,1,1)$, we conclude that $\P(\Er_2) \to 0$ if
\begin{align*}
	R_1 + \min\{ R_2, H(X_{21}\cond  X_{31},Q) \}&< I(X_1,X_{21}; Y_1 \cond  X_{31},Q),
\end{align*}
or, equivalently,
\begin{align}
	R_1 + \min\{ R_2, H(X_{21}\cond Q) \} &< H(Y_1\cond  X_{31},Q).
\end{align}
$\P(\Er_3)$ can be analyzed in an identical fashion to $\P(\Er_2)$, and we have $\P(\Er_3) \to 0$  as $n \to \infty$ if
\begin{align}
	R_1 + \min\{ R_3, H(X_{31}\cond Q) \} &< H(Y_1\cond  X_{21},Q).
\end{align}
Finally, the event $\Er_4$ is augmented as 
\begin{align*}
	\Er_4 &\subseteq \Bigl\{ \left( Q^n, X_1^n(m),  S^n_1(l,k), Y_1^n(1,1,1) \right)  \in \Typ \\
	& \qquad \text{ for some } m,l,k\neq 1 \Bigr \} .
\end{align*}
Lemma~\ref{lemma:spl} with $U^n=Q^n$, $A^n=X_1^n$, $B_1^n=X_{21}^n$, $B_2^n=X_{31}^n$, $B^n=S_1^n$, $h=h_1$, and $C^n=Y_1^n(1,1,1)$ shows that $\P(\Er_4)\to 0$ as $n \to \infty$ if
\begin{align}
	R_1 + \min\{ R_2+R_3, R_2+H(X_{31}\cond Q), \ \, \notag \\
	H(X_{21}\cond Q)+R_3, H(S_1\cond Q) \} &< H(Y_1\cond Q) , \label{eq:lastcond2}
\end{align}
where we have used $I(X_1, S_1; Y_1\cond Q)=H(Y_1\cond Q)$.
Collecting~\eqref{eq:firstcond2} to~\eqref{eq:lastcond2} yields the conditions of $\Rr_1$. The probability of error at the second and third receiver can be bounded similarly, leading to the conditions of $\Rr_2$ and $\Rr_3$. Finally, the cardinality bound on $\Qc$ can be established using the bounding technique described in~\cite{ElGamalKimLecture}. \hfill \IEEEQED

\section{Proof of Theorem~\ref{thm:ID_is_capacity}}  \label{sec:proof_ID_is_capacity}
\begin{IEEEproof}[Proof of achievability]
	We prove achievability with interference decoding. Specifically, we show that under strong interference and invertible $h_k$, regions $\Rr_k$ of Theorem~\ref{thm:ID} simplify to regions $\Rr''_k$ below while maintaining $\Rr_1 \cap \Rr_2 \cap \Rr_3 = \Rr''_1 \cap \Rr''_2 \cap \Rr''_3$.
	
	Recall the definition of $\Rr_1(Q,X_1,X_2,X_3)$ as the set of rate triples $(R_1,R_2,R_3)$ that satisfy inequalities~\eqref{eq:IDcond1} to~\eqref{eq:IDcond4}.
	Further recall the analogous definitions of $\Rr_2$ and $\Rr_3$, which include the inequalities
		\begin{align}
			R_2 &< H(X_{22} \cond Q), \label{eq:r2_1} \\
			R_3 &< H(X_{33} \cond Q). \label{eq:r3_1} 
		\end{align}
	When combined with the strong interference assumption, inequalities~\eqref{eq:r2_1} and~\eqref{eq:r3_1} imply that the $\min$ expressions in~\eqref{eq:IDcond2} and~\eqref{eq:IDcond3} simplify to $R_2$ and $R_3$, respectively. Furthermore, the sum of~\eqref{eq:r2_1} and~\eqref{eq:r3_1} implies that
		\begin{align*}
			R_2 + R_3 &< H(X_{22} \cond Q) +  H(X_{33} \cond Q) \\
			&\leq H(X_{21} \cond Q) +  H(X_{31} \cond Q) \\
			&= H(S_1 \cond Q),
		\end{align*}
	where we have used the invertibility of $h_1$. Therefore, the $\min$ expression in~\eqref{eq:IDcond4} simplifies to $R_2+R_3$. 

	Consequently, define $\Rr''_1(Q,X_1,X_2,X_3)$ as the set of rate triples $(R_1,R_2,R_3)$ such that 
		\begin{align*}
				R_1 &< H(X_{11} \cond Q),   \\
				R_1 + R_2 &< H(Y_1\cond  X_{31},Q),  \\
				R_1 + R_3 &< H(Y_1\cond  X_{21},Q),  \\
				R_1 + R_2+R_3 &<  H(Y_1\cond Q). 
		\end{align*}
	Likewise, define the regions $\Rr''_2(Q,X_1,X_2,X_3)$ and $\Rr''_3(Q,X_1,X_2,X_3)$ by replacing subscripts following $1\mapsto 2 \mapsto 3 \mapsto 1$  and $1\mapsto 3 \mapsto 2 \mapsto 1$ in $\Rr''_1(Q,X_1,X_2,X_3)$, respectively. Then Theorem~\ref{thm:ID} implies that 
			$$\bigcup_{(Q,X_1,X_2,X_3)} \bigcap_{k=1}^3 \Rr''_k(Q,X_1,X_2,X_3)$$
	is achievable, and the proposition follows by expanding the intersection operation.
\end{IEEEproof}

\begin{IEEEproof}[Proof of converse]
	Consider a code with rates $(R_1,R_2,R_3)$, empirical pmf $p(x_1^n)p(x_2^n)p(x_3^n)$, and  $\pen$ tending to $0$ as $n \to \infty$.
	First, note that
	\begin{align*}
		nR_1 &\leq I(X_1^n; Y_1^n) + n\eps_n \\
		&=I(X_{11}^n; Y_1^n) + n\eps_n \\
		&\leq H(X_{11}^n)+ n\eps_n \\
		&= nH(X_{11}\cond Q)+ n\eps_n,
	\end{align*}
	where $Q$ is a time-sharing random variable uniformly distributed over $[1:n]$. Next, consider
	\begin{align*}
		& n(R_1+R_2) \\
		&\leq I(X_1^n; Y_1^n) + I(X_2^n; Y_2^n) + n\eps_n \\
		&= H(Y_1^n) - H(Y_1^n \cond X_1^n) + H(Y_2^n) - H(Y_2^n \cond X_2^n)  + n\eps_n \\
		&= H(Y_1^n) - H(S_1^n) + H(Y_2^n) - H(S_2^n) + n\eps_n \\
		&= H(Y_1^n) - H(X_{31}^n) \\
		&\quad + \left( H(Y_2^n) - H(X_{21}^n)   - H(X_{12}^n) - H(X_{32}^n) \right) + n\eps_n \\
		&\leq H(Y_1^n \cond X_{31}^n) + n\eps_n \\
		&\leq  nH(Y_1 \cond X_{31}, Q) + n\eps_n,
	\end{align*}
	where we have used $H(X_{22}^n) \leq H(X_{21}^n)$ and $H(Y_2^n) \leq H(X_{22}^n)  + H(X_{12}^n) + H(X_{32}^n)$.
	In the same way, it can be shown that 
	\begin{align*}
		n(R_1+R_3) &\leq nH(Y_1 \cond X_{21}, Q) + n\eps_n.
	\end{align*}
	Finally, 
	\begin{align*}
		&n(R_1+R_2+R_3) \\
		&\leq H(Y_1^n) - H(S_1^n) + H(Y_2^n) - H(S_2^n) \\
		&\quad +  H(Y_3^n) - H(S_3^n)  + n\eps_n \\
		&= H(Y_1^n) + n\eps_n \\
		&\quad + \left( H(Y_2^n) - H(X_{21}^n) - H(X_{12}^n) - H(X_{32}^n)   \right) \\
		&\quad + \left( H(Y_3^n) - H(X_{31}^n) - H(X_{13}^n) - H(X_{23}^n)    \right) \\
		&\leq nH(Y_1 \cond Q) + n\eps_n.
	\end{align*}
	Thus, all four conditions related to the first receiver have been shown. Analogous steps yield the remaining bounds. 
\end{IEEEproof}

\section{Proof of Theorem~\ref{thm:inclusion}}  \label{sec:proof_inclusion}
We show that the inner bound in Lemma~\ref{lemma:tian} is included in the inner bound of Theorem~\ref{thm:ID}.
The conditions of region $\Rr_1$ in Theorem~\ref{thm:ID} can be made more stringent by replacing the $\min$ expression with any one of its argument terms. For example, $(R_1,R_2,R_3)\in \Rr_1$ is implied by
\begin{align*}
			R_1 &< H(X_{11} \cond Q), \\
			R_1 + H(X_{21}\cond Q) &< H(Y_1\cond  X_{31},Q),  \\
			R_1 + H(X_{31}\cond Q) &< H(Y_1\cond  X_{21},Q),  \\
			R_1 + H(S_1\cond Q) &<  H(Y_1\cond Q),
\end{align*}
or, equivalently,
\begin{align}
	R_1  &< \min \{
		H(X_{11}| Q),  \notag \\
		&\qquad \quad\ \,  H(Y_1|  X_{31},Q) - H(X_{21}| Q),  \notag \\
		&\qquad \quad\ \,  H(Y_1|  X_{21},Q)  -  H(X_{31}| Q), \notag \\
		&\qquad \quad\ \,  H(Y_1| Q)  -  H(S_1| Q)
	\}. \label{eq:minoffour}
\end{align}
To simplify this expression, consider
\begin{align*}
	H(X_{11}\cond Q) &\geq I(X_{11}; Y_1\cond Q) \\
	& = H(Y_1\cond Q) - H(Y_1\cond X_{11},Q) \\
	&= H(Y_1\cond Q) - H(S_1\cond Q),
\end{align*}
as well as
\begin{align*}
	&\left[ H(Y_1\cond X_{21},Q) - H(X_{31}\cond Q) \right] - \left[ H(Y_1\cond Q)-H(S_1\cond Q) \right] \\
	&\qquad = H(Y_1, X_{21}\cond Q) - H(X_{21}\cond Q)  \\
	&\qquad \qquad  - \underbrace{H(X_{31}\cond Q)}_{H(S_1\cond X_{21},Q)}  - H(Y_1\cond Q) + H(S_1\cond Q) \\
	&\qquad = H(X_{21}\cond Y_1,Q) - H(X_{21}\cond S_1,Q)\\
	&\qquad \geq H(X_{21}\cond Y_1, S_1,Q) - H(X_{21}\cond S_1,Q) \\
	&\qquad = 0,
\end{align*}
and, by symmetry,
\begin{align*}
	\left[ H(Y_1| X_{31},Q)\!-\!H(X_{21}| Q) \right] - \left[ H(Y_1| Q)\!-\!H(S_1| Q) \right]
	&\geq  0.
\end{align*}
Thus, the $\min$ in~\eqref{eq:minoffour} is always achieved by the last term, and~\eqref{eq:minoffour} simplifies to 
\begin{align*}
	R_1 &< H(Y_1\cond Q) - H(S_1\cond Q) \\
	&= I(X_1; Y_1\cond Q).
\end{align*}
Using a similar argument, it follows that the conditions for $\Rr_2$ and $\Rr_3$ in Theorem~\ref{thm:ID} are implied by~\eqref{eq:tian}.  \hfill \IEEEQED

\begin{remark}
In the case with noisy observations, this proof fails in the following manner. Interference decoding entails the inequality
\begin{align*}
	R_1 &< I(X_1,S_1; Z_1 \cond Q) - H(S_1\cond Q) \\
	&= I(X_1; Z_1 \cond Q) + I(S_1; Z_1 \cond X_1, Q) - H(S_1\cond Q) \\
	&= I(X_1; Z_1 \cond Q) + H(S_1 \cond X_1, Q) - H(S_1\cond X_1, Z_1, Q) \\
	&\quad - H(S_1\cond Q) \\
	&= I(X_1; Z_1 \cond Q) - H(S_1\cond X_1, Z_1, Q) .
\end{align*}
The first term is the achievable rate when treating interference as noise. The second term is zero when the channel is noiseless and acts as a penalty when noise is introduced.
\end{remark}

\section{Final remarks} \label{sec:finalRemarks}
This paper presented an interference-decoding inner bound to the capacity region of  a class of three-user-pair deterministic interference channels. We showed that this inner bound strictly includes the interference-as-noise region. 
As in treating interference as noise, the interference-decoding scheme uses point-to-point codes. The  decoder in interference decoding, however, is more sophisticated. 

We showed that interference decoding is optimal under strong interference and function invertibility conditions. The scheme is not optimal in general, however. To exemplify this, we consider the following two-user-pair deterministic interference channel for which the capacity region is known.  

\begin{figure}[t!]
	\centering
	\subfigure[Block diagram of the channel.]{
		\includegraphics{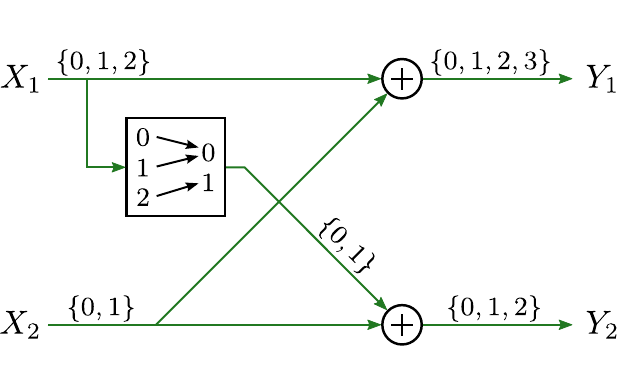}
		\label{fig:example2dic}
	}
	\hfill
	\subfigure[Capacity region and inner bounds.]{
		\includegraphics{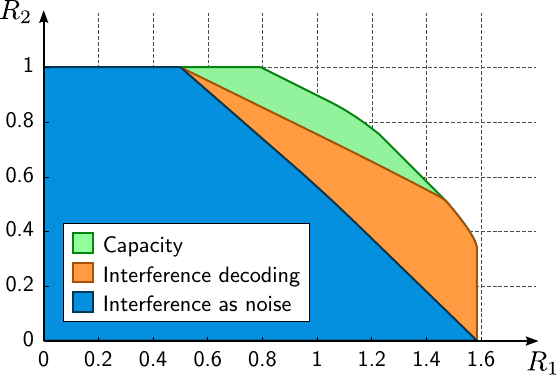} 
		\label{fig:ElG_Costa_nonsym_modified_recolored}
	}
	\caption{2-DIC example.}
\end{figure}

Consider the 2-DIC in Figure~\ref{fig:example2dic} with input alphabets $\mathcal X_1=\{0,1,2\}$, $\mathcal X_2=\{0,1\}$, loss functions $g_{12} = \{0 \mapsto 0, 1\mapsto 0, 2 \mapsto 1\}$ and $g_{11} = g_{22}=g_{21}=\mathrm{Id}$, and receiver functions $f_1=f_2$ being addition. (The interference combining functions $h_1$ and $h_2$ are not relevant in this case.) 
The outputs of the channel are thus given by
\begin{align*}
	Y_{1} &= X_{1} + X_{2}, \\
	Y_{2} &= g_{12}(X_{1}) + X_{2}.
\end{align*}

The interference-decoding inner bound in Theorem~\ref{thm:ID} reduces to the set of rate pairs $(R_1,R_2)$ that such that
\begin{align*}
	R_1 &< H(X_1 \cond Q), \\
	R_2 &< H(X_2 \cond Q), \\
	R_1 + \min\{R_2, H(S_1 \cond Q) \} &< H(Y_1 \cond Q), \\
	R_2 + \min\{R_1, H(S_2 \cond Q) \} &< H(Y_2 \cond Q),
\end{align*}
for some $p(q) p(x_1|q) p(x_2|q)$. 

Figure~\ref{fig:ElG_Costa_nonsym_modified_recolored} compares this inner bound to the capacity region given in~\cite{ElGamalCosta82} and to the region achievable by treating interference as noise (Lemma~\ref{lemma:tian}). Not surprisingly, interference decoding does not achieve the full capacity. To achieve capacity, Han--Kobayashi rate splitting and superposition coding are needed.

\appendix

\subsection{Proof of Lemma~\ref{lemma:mpl}}
Applying the packing lemma~\cite{ElGamalKimLecture} 
with $U = U$, $X = (A,B)$, and $Y = C$ immediately establishes the convergence if $R_A+R_B< I(A,B;C \cond  U) - \delta(\varepsilon)$. Next, we prove convergence when $R_B + H(A\cond U) < I(A,B;C \cond  U)-\delta(\eps)$. To this end, we bound the probability of the event in question as follows
	\begin{align*}
		& \P \{ (U^n,A^n(m),B^n(l), C^n)\in \Typ \text{ for some } m,l \} \notag \\
		& \leq \sum_{l=1}^{2^{nR_B}} \P \{ (U^n,A^n(m),B^n(l), C^n)\in \Typ \text{ for some } m\} \notag 
	\end{align*} 
	\begin{align*}
		& = \sum_{l=1}^{2^{nR_B}}  \sum_{u^n \in \Typ(U)}  \P(U^n=u^n) \notag \\
		& \qquad \cdot \hspace{-6mm} \sum_{b^n \in \Typ(B\cond u^n)} \P \{ B^n(l)=b^n \cond  U^n=u^n\} \notag \\
		& \qquad \cdot \hspace{-9mm} \sum_{\substack{
			a^n(m) \in \Typ(A\cond u^n)  \notag \\
			\text{ for all } m \in [1:2^{nR_A}]}}
		\P\{ A^n(m)=a^n(m) \text{ for all } m \cond  U^n=u^n\} \\
		& \qquad \cdot \P \left\{ \bigcup_{m=1}^{2^{nR_A}} \{ (u^n,a^n(m),b^n, C^n)\in \Typ(U,A,B,C) \} \right\}.
	\end{align*}
Upon closer inspection, the union in the last probability term potentially contains duplicate events, for example if $a^n(1)=a^n(2)$. Those duplicates can be eliminated. By pessimistically assuming that the set $\{a^n(m):\, m \in [1:2^{nR_A}]\}$ is equal to $\Typ(A\cond u^n)$, we can write 
	\begin{align}
		& \P\left \{ \bigcup_{m=1}^{2^{nR_A}}  \{ (u^n,a^n(m),b^n, C^n)\in \Typ(U,A,B,C)  \} \right \} \notag \\
		&\leq \P \left\{ \bigcup_{a^n \in \Typ(A\cond u^n)} \hspace{-6mm} \{ (u^n,a^n,b^n, C^n)\in \Typ(U,A,B,C) \} \right\} \notag \\
		&\leq \sum_{a^n \in \Typ(A\cond u^n)} \P\{ (u^n,a^n,b^n, C^n)\in \Typ(U,A,B,C)  \} \notag \\
		&\annleq{a}  2^{n(H(A\cond U)+\delta_1(\eps))} \cdot 2^{-n(I(C;A,B\cond  U)-\delta_2(\eps))} \notag \\
		&= 2^{n(H(A\cond U) - I(C;A,B\cond  U) +\delta(\eps))}. 
		\label{eq:lemma1_proof_2}
	\end{align}
In step (a), we use the upper bound on the size of the conditional typical set $\Typ(A\cond u^n)$ and the joint typicality lemma~\cite{ElGamalKimLecture} 
with $X=U$, $Y=(A,B)$ and $Z=C$. Strictly speaking, the joint typicality lemma holds only if $(u^n,a^n,b^n) \in \Typ(U,A,B)$, which does not necessarily follow from $(u^n,a^n) \in \Typ(U,A)$ and $(u^n,b^n) \in \Typ(U,B)$ as given in our sum expression. However, for the cases where $(u^n,a^n,b^n) \notin \Typ(U,A,B)$, we have $\P \{ (u^n,a^n,b^n, C^n)\in \Typ(U,A,B,C) \}=0$, and the bound from the joint typicality lemma still holds (though very loosely). Substituting from \eqref{eq:lemma1_proof_2} into  
the previous inequality, we obtain
	\begin{align*}
		& \P \left \{ (U^n,A^n(m),B^n(l), C^n)\in \Typ \text{ for some } m,l \right \} \\
		& \leq 2^{n(R_B+ H(A\cond U) - I(C;A,B\cond  U) +\delta(\eps))}.
	\end{align*}
Clearly, this probability converges to zero as $n\to \infty$ if $R_B + H(A\cond U) < I(A,B;C\cond U)-\delta(\eps)$.
Completely symmetrically, convergence follows from $R_A + H(B\cond U) < I(A,B;C\cond U)-\delta(\eps)$.
Thus convergence is implied by $\min \left \{R_A+R_B, R_A+H(B\cond U), H(A\cond U)+R_B \right \} < I(A,B;C \cond  U) -\delta(\eps)$, and the desired result follows by recalling that $H(A \cond U)+H(B \cond U)\geq I(A,B;C \cond U)$. \hfill \IEEEQED

\subsection{Proof of Lemma~\ref{lemma:spl}}
The first and last term in the $\min$ expression follow immediately from Lemma~\ref{lemma:mpl} by disregarding the special structure of $B^n$. For the second term, we argue similarly,
	\begin{align*}
		& \P \left\{ (U^n,A^n(m),B^n(l_1,l_2), C^n)\in \Typ \text{ for some } m,l_1,l_2 \right\} \notag \\
		& \leq  \sum_{u^n \in \Typ(U)}  \P(U^n=u^n) 
		\sum_{m=1}^{2^{nR_A}} \notag \\
		& \quad \qquad \cdot \hspace{-6mm} \sum_{a^n \in \Typ(A\cond u^n)} \hspace{-5mm} \P \left(A^n(m)=a^n \cond  U^n=u^n\right)   \notag \\
		& \quad \qquad \cdot \sum_{l_1=1}^{2^{nR_{B_1}}}
		\sum_{b_1^n \in \Typ(B_1\cond u^n)} \hspace{-6mm} \P\left(B_1^n(l_1)=b_1^n \cond  U^n=u^n\right) \notag \\
		& \quad \qquad \cdot \hspace{-9mm} \sum_{\substack{
			b_2^n(l_2) \in \Typ(B_2\cond u^n)\\
			\text{for all } l_2 \in [1:2^{nR_{B_2}}]}} 
		\hspace{-9mm} \P\left(B_2^n(l_2)=b_2^n(l_2) \text{ for all $l_2$} \cond  U^n=u^n\right)   \notag \\
		& \quad \qquad \cdot \P \left\{ \bigcup_{l_2=1}^{2^{nR_{B_2}}} \left \{ (u^n,a^n,b^n(b_1^n,b_2^n(l_2)), C^n)\in \Typ \right \} \right \}.
	\end{align*}
There are at most $\lvert \Typ(B_2\cond u^n) \rvert $ distinct events in the union expression. Using a similar line of reasoning as in the proof of Lemma~\ref{lemma:mpl}, we can upper bound the last probability term by $2^{n(H(B_2\cond U) - I(C;A,B\cond  U) +\delta(\eps))} $.
Substituting into 
the previous inequality, we obtain the upper bound $2^{n(R_A+R_{B_1}+ H(B_2\cond U) - I(C;A,B\cond  U) +\delta(\eps))}$ on the probability of the event of interest. Clearly, this expression converges to zero as $n\to \infty$ if $R_A + R_{B_1} + H(B_2\cond U) < I(A,B;C\cond U)-\delta(\eps)$. We have thus established the second term in the $\min$ expression. The third term follows in a symmetric manner.\hfill \IEEEQED

\bibliographystyle{IEEEtran}
\bibliography{IEEEabrv,references}

\end{document}

%% file: defns.tex
\usepackage{xspace}
\usepackage{mathrsfs}
\usepackage{amsopn}

\newcommand{\Qc}{\mathcal{Q}}

\newcommand{\Xc}{\mathcal{X}}
\newcommand{\Yc}{\mathcal{Y}}

\newcommand{\Cr}{\mathscr{C}}

\newcommand{\Rr}{\mathscr{R}}

\newcommand{\pen}{{P_e^{(n)}}}

\def\eps{\epsilon}

\let\P\relax
\DeclareMathOperator\P{P}

\def\textiid{i.i.d.\@\xspace}
\newcommand\iid{\ifmmode\text{ i.i.d. } \else \textiid \fi}

\newcommand{\Real}{\mathbb{R}}
